\documentstyle[12pt,aasms4]{article}

\slugcomment{To appear in Astrophysical Journal Letters}

\lefthead{Burleigh, Jordan \and Schweizer}
\righthead{FUV HST Observations of RE~J0317$-$853}

\begin{document}

\title{Phase-resolved far-ultraviolet HST spectroscopy of the
peculiar magnetic white dwarf RE J0317$-$853}

\author{M. R. Burleigh}
\affil{Department of Physics and Astronomy, University of
                Leicester, Leicester LE1 7RH, UK}

\author{S. Jordan}
\affil{Institut f\"ur Astronomie und Astrophysik der
Universit\"at, D-24098 Kiel, Germany}

\and

\author{W. Schweizer}
\affil{Institut f\"ur Astronomie und Astrophysik, Auf der Morgenstelle 10, 
D-72076 T\"ubingen, Germany}



\begin{abstract}

We present phase resolved FUV HST FOS spectra of the rapidly
rotating, highly magnetic white dwarf RE J0317$-$853. Using these data,
we construct a new model for the magnetic field morphology 
across the stellar surface. From an expansion into spherical harmonics,
we find the range of magnetic field strengths present is 180$-$800MG. 
For the first time we could 
identify an absorption feature present at certain phases 
at $\approx$1160{\AA} as a ``forbidden'' $1s_0 \rightarrow 2s_0$ 
component, due to the combined presence of an  electric and magnetic
field.

\end{abstract}




%

\section{Introduction}

RE~J0317$-$853 is a 
highly unusual magnetic DA white dwarf, discovered
by Barstow et~al. (1995, hereafter B95) during the optical follow-up
campaign to the ROSAT Wide Field Camera (WFC) all sky survey (Pounds
et~al., 1993). B95 compared the optical spectrum and a far-UV (FUV)
spectrum obtained with IUE to synthetic spectra generated for a number of
different field configurations, and found that it was best
matched by a dipole model, with a polar field strength of $\approx$340\,MG,
but which is offset along the dipole axis by 0.2 stellar radii in the
direction of the southern magnetic pole.  In this model, the range of field
strengths seen by an observer
varies between $\approx$660\,MG at the southern pole 
to $\approx$200\,MG at the 
north pole. A later analysis of 
circular polarisation data by Ferrario
et~al. (1997, hereafter F97) produced a broadly similar best-fit model:
$B=450$\,MG, displaced by 35\% of the stellar radius and viewed at an angle
of $30^\circ-60^\circ$ to the dipole axis.

Fixing the gravity at log g$=$8.0, the temperature was found, from the
shape and slope of the continuum in the FUV spectrum, to be
$49,250_{-1,200}^{+2,150}$\,K. Thus, not only is RE J0317$-$853 one of the
most magnetic white dwarfs ever found, it is the hottest member 
of this subclass of degenerate objects.

This unique star set several other records for white dwarfs. 
The close proximity of an ordinary DA white dwarf, LB09802, just
$16^{\prime \prime}$ away, allowed B95 to estimate the mass of RE J0317$-$853 at
1.35M$_\odot$, approaching the Chandrasekhar limit. 
In addition, the discovery that it was rapidly and
regularly varying in the optical led B95 to conclude that it must be
rotating with a period of just 725 seconds, making RE J0317$-$853 the
most rapidly rotating isolated white dwarf ever found. Both the 
fast rotation and the high mass may be an indication that RE J0317$-$853
is the end product of a double degenerate merger.

Comparing models to the flux spectrum of a highly magnetic white dwarf can 
yield a variety of good, but not necessarily unique fits (Putney and
Jordan, 1995). 
The optical and FUV flux spectra obtained  by B95 
for RE J0317$-$853 give
the field strength over the whole star averaged out over several
rotations. B95 had no direct information about, or a unique solution to,
either the direction of the field or the orientation of the magnetic
axis. Although F97 did attempt to add the circular polarisation spectra in
phase, they had no knowledge prior to their observations of the optical
period, and could not provide a detailed model of the field structure
across the whole stellar surface.

In addition, the model described above does not fit the FUV IUE
spectrum in detail. Here, 
in contrast to the complicated structure of the optical spectrum where
the Balmer lines are shifted almost beyond recognition by thousands of
{\AA}, the mean FUV IUE spectrum is comparatively simple. Lyman
$\alpha$ is split, with the redward shifted 
$\sigma$ ($1s_0 \rightarrow 2p_{+1}$, in the field free notation)
 component being
visible in the IUE SWP spectrum. The blueward shifted $\sigma$ component
lies outside the range of this instrument. However, only one redward
Lyman $\alpha$ component is predicted by the B95 and F97 models (at
1300{\AA}), yet a second deep 
absorption features is present near 1340{\AA}. Such a  doublet structure could
be explained if we are seeing two distinct regions with rather different
field strengths. 

This hypothesis could only be tested with the help of phase resolved
spectroscopy, by which --- depending on the angle between the observer and
the rotational axis  --- 
a major part of the global field can be mapped.
A `true' field configuration is one that, when seen at different
angles due to the rotational phase, matches the data at each observation. 
In this paper we present time-resolved FUV spectra
of RE J0317$-$853  obtained
with the Faint Object Spectrograph (FOS) of the 
Hubble Space Telescope (HST)
to look for variations in the line positions and strengths of the 
Lyman $\alpha$ 
components, and try to determine  the field morphology across the
stellar surface by applying radiative transfer calculations.

\section{HST Observations and Data Reduction}

FUV observations were obtained across two orbits with FOS on 
1996 November 26, using  the blue filter, G130H grating and the 
$1^{\prime \prime}$ 
 aperture, giving approximately 1{\AA} resolution. A total of 142 
$\times$35 second exposures were obtained in RAPID mode, 58 on the first 
orbit and 84 on the second. The use of RAPID mode allowed us to minimise 
the readout time between exposures, and maximise the phase coverage across 
the 725 second rotation period. The total exposure time was 4858.6 
seconds. The data were extracted with standard IRAF/STSDAS routines, and 
binned into 12 equal phase intervals (Fig.\,1). 
Distinct changes in the FUV spectrum are clearly 
visible across the rotation period. In particular, the position of the 
redward $\sigma$ Lyman $\alpha$ component
appears to shift  
from $\approx$1340{\AA} at phase 0.167 (phase=0 is defined by our first
bin during the HST orbits) to $\approx$1300{\AA} at 
phase 0.667,  
while the  $\pi$ ($1s_0 \rightarrow 2p_0$)  
Lyman $\alpha$ component is also seen to grow in strength to a maximum at 
phase 0.667. One other intriguing absorption feature at  $\approx$1160{\AA}
is seen to grow in strength at later 
phases, in conjunction with the $\approx$1300{\AA} and stationary Lyman
$\alpha$ lines. 

\section{Analysis and Results}

The observed spectrum and wavelength dependent polarisation of magnetic white
dwarfs, described  by the four Stokes parameters, 
can be analysed by simulating the transport of polarized radiation
through a magnetized stellar atmosphere
(e.g. Jordan, 1992, Putney \&\ Jordan, 1995). However, in 
previous analyses it
was extremely time consuming to calculate new synthetic spectra for
every variation of the magnetic field geometry. In particular, when it comes
to  phase resolved spectroscopy, a different approach is necessary, 
which we have adopted for the first time here.

We calculated a library of synthetic spectra for a given effective
temperature (40\,000\,K), for different field strengths
between 100\, and 900\,MG (in steps of 1\,MG) and for nine different
angles (equidistant in the cosine) between the magnetic field 
and the observer. Limb darkening was taken into account by a linear
approximation based on an interpolation between model spectra: 
$I(\mu)= (0.7+0.3 \mu) I(\mu=1)$ (Euchner, 1998). 

In describing the magnetic field geometry we utilised two different
approaches: (i) pure dipole models with the ability to 
offset the dipole 
center relative to the center of the star, and (ii) a   
more general approach
involving an  expansion into spherical harmonics. 

For a clearly rotating magnetic star, different 
spectra are observed depending on the relative contribution of the magnetic
field, and the angles between the observer and the magnetic field at a given
phase. After prescribing a specific global magnetic field, three
additional free parameters are needed to calculate a synthetic spectrum
for a given phase: (i) the angle $90^{\circ}-i$  between  the rotational axis 
and the
observer, (ii) the angle $\beta$  between the rotational axis and the magnetic 
($z-$) axis (relative to which the coordinate system of the
offset dipoles or the spherical harmonics have been defined),
and (iii) the phase at which the magnetic pole passes the meridian
between the rotational axis and the center of the visible hemisphere.

The ``best'' model is found by minimizing the $\Delta \chi^2$ 
between the predicted flux and the observations for all the 
different phases into which the HST observations have been binned.
The solution is then found by varying the free parameters in a 
systematic way, using the ``Amoeba'' 
 downhill simplex method in multidimensions (Nelder \& Mead, 1965, 
Press et al., 1986). 

In our current analysis we did not vary the effective temperature,
although we know from the optical ($\Delta m_V=\pm 0\fm 1$, see B95)
and our UV data ($\Delta m_{\mbox{\scriptsize 1500\,\AA}}=\pm 0\fm 1$) 
that the surface brightness is changing 
with the rotational phase. This effect may be due to the
influence of magnetic pressure on the 
structure of the atmosphere, but has  not  been taken into account here.
Since the fit was slightly better with $T_{\rm eff}=40000$\,K than
with 30000 or 50000\,K we fixed that value and adjusted the theoretical
and observed flux to the same value at 1500\,\AA.

Our first attempt was to fit the data with a centered dipole. We were
able to  roughly fit the data 
with a polar  field strength of 220\,MG  at phase=0.167 
or with  450\,MG at phase=0.667, 
but not for all phases simultaneously.
The next step was the introduction of an offset along the magnetic
$z$ axis. A reasonable fit was found for a dipole field
strength of 363\,MG, $z_{\rm off}=-0.19$\,stellar radii, $i=39^{\circ}$
and $\beta=20^{\circ}$, very consistent
with the solution inferred from the (phase averaged) discovery spectrum
in B95. 

A slightly better solution was obtained by additional offsets perpendicular
to the $z$ axis (see Fig.\,1a in Putney \&\ Jordan, 1995): $x_{\rm off}=0.057$,
$y_{\rm off}=-0.04$, $z_{\rm off}=-0.220$. In this case the angles
were different, resulting in $i=40^{\circ}$ and $\beta=42^{\circ}$
(leading to a visible range of surface field strengths between 140
and 730\,MG);
the differences caused by the perpendicular offsets were partly
compensated by the $\beta$ angle. Since the fit was not perfect in
either case 
(reduced $\chi^2\approx 4$ and $3$, respectively), we could conclude that
there is no contradiction to the assumption of a purely axis-symmetrical field.

This conclusion could also be drawn from an expansion of the surface
magnetic fields into:

\begin{displaymath}
\begin{array}{lll}
B_r=&-\sum_{l=1}^\infty \sum_{m=0}^l (l+1) (g_l^m \cos m\phi
+h_l^m \sin m \phi)\\ &\hfill  P_l^m(\cos\theta)\\
B_\theta=+ &\sum_{l=1}^\infty \sum_{m=0}^l  (g_l^m \cos m\phi
+ h_l^m \sin m \phi)\\ & \hfill dP_l^m(\cos\theta)/d\theta\\
B_\phi=&-\sum_{l=1}^\infty \sum_{m=0}^l  m (g_l^m \cos m\phi
+ h_l^m \sin m \phi)\\&\hfill dP_l^m(\cos\theta)/d\sin \theta
\end{array}
\end{displaymath}
with the associated Legendre polynomials $P_l^m$.

For this paper we limited ourselves to $l\le 3$,
i.e. to 15 free parameters. The twelve parameters
with $0<m<l$, and therefore $B_\phi$,  were very small ($<10$\,MG) 
compared to the axial symmetric
contributions from $m=0$ ($g_1^0=-206$\,MG , $g_2^0=+77$\,MG,
$g_3^0=-39$\,MG). 
Possibly, slightly  better fits can be obtained by
including $l>3$. However,  this does not yet seem to be reasonable
since we can expect systematic errors introduced by
our assumptions (e.g. constant effective temperature). 

In this model (with $\beta=29^{\circ}$ and $i=34^{\circ}$)
 the field strengths cover the range 180-800\,MG.
However, it is not quite clear whether fields up to 800\,MG
are indeed present, since the the $1s_0 \rightarrow 2p_{+1}$
component has a maximum (1343\,\AA) at 470\,MG, so that  higher
and lower field strengths lead to the same shift. 
However, both the B95 and F97  analyses  of the optical and
polarimetric data indicate rather strong fields
(200-660\,MG, and 180-1600\,MG, respectively).

Qualitatively, the bimodal behaviour of  RE J0317$-$853 can be understood
in the following way: during the ``low field'' phases (when $B<210$\,MG 
dominates,
see Fig.\,4) the $1s_0 \rightarrow 2p_0$ and 
$1s_0 \rightarrow 2p_{+1}$ 
components are  centered at about  1195\,\AA\ and 1303\,\AA,
respectively. When the ``high'' magnetic fields ($B>280$\,MG) become more
important, the $1s_0 \rightarrow 2p_0$ is shifted to smaller wavelengths,
out of the HST range, and $1s_0 \rightarrow 2p_{+1}$ is close to its
maximum at 1343\,\AA. 

One unexpected feature, not recognized in the IUE spectrum, 
is an additional absorption line 
at $\approx 1160$\,\AA. This appears during the ``low field'' phase,  whenever
 $1s_0 \rightarrow 2p_0$ is visible as well. 
In the dense atmosphere of magnetic white dwarf stars, mean electric 
field strengths up to $10^7 $ V/m can be induced by free electrons and
ions. The hydrogen atom can experience, due to thermal motion, an additional 
Lorentzian electric field up to $10 ^7 \cdots 10^9$ V/m, perpendicular to the
magnetic field  (Mathys, 1989). This gives rise to the motional Stark effect.
Without additional electric fields, the magnetic quantum number m and the 
z-parity
$\pi_z$ are conserved. For parallel electric and magnetic fields, only 
the magnetic quantum number $m$ is conserved, 
and for perpendicular electric and 
magnetic fields only the $z$-parity. In contrast, no discrete symmetry is
left in arbitrarily oriented fields. 
Therefore, in combined electric and magnetic fields, 
additional allowed dipole transitions occur (e.g. $1s_0 \rightarrow 2s_0$,
which gains oscillator strength at the expense of the
 $1s_0 \rightarrow 2p_0$)  compared to the diamagnetic system, where only
three transitions ($1s_0 \rightarrow 2p_{-1,0,+1}$) are dipole allowed.
Recognizing this possibility, 
we included the ``forbidden'' $1s_0 \rightarrow 2s_0$
in our calculations and, consequently, 
could reproduce the peculiar feature found at 1160\,\AA.

Our model for the magnetic field is based on the three
visible Lyman $\alpha$ components (including the ``forbidden'' one) 
only. As we have seen, different
assumptions about the field geometry lead to  fits of similar
quality. Although the basic properties of the offset dipole model and the
expansion into spherical harmonics are the same we expect that we 
will be able to constrain the global magnetic field much better when we
include, in the near  future, phase resolved spectropolarimetric data.

\acknowledgments

We acknowledge the support of PPARC, UK,
DFG (Ko 738/7-1) and DLR (50 OR 97042).

\clearpage

\clearpage

\figcaption[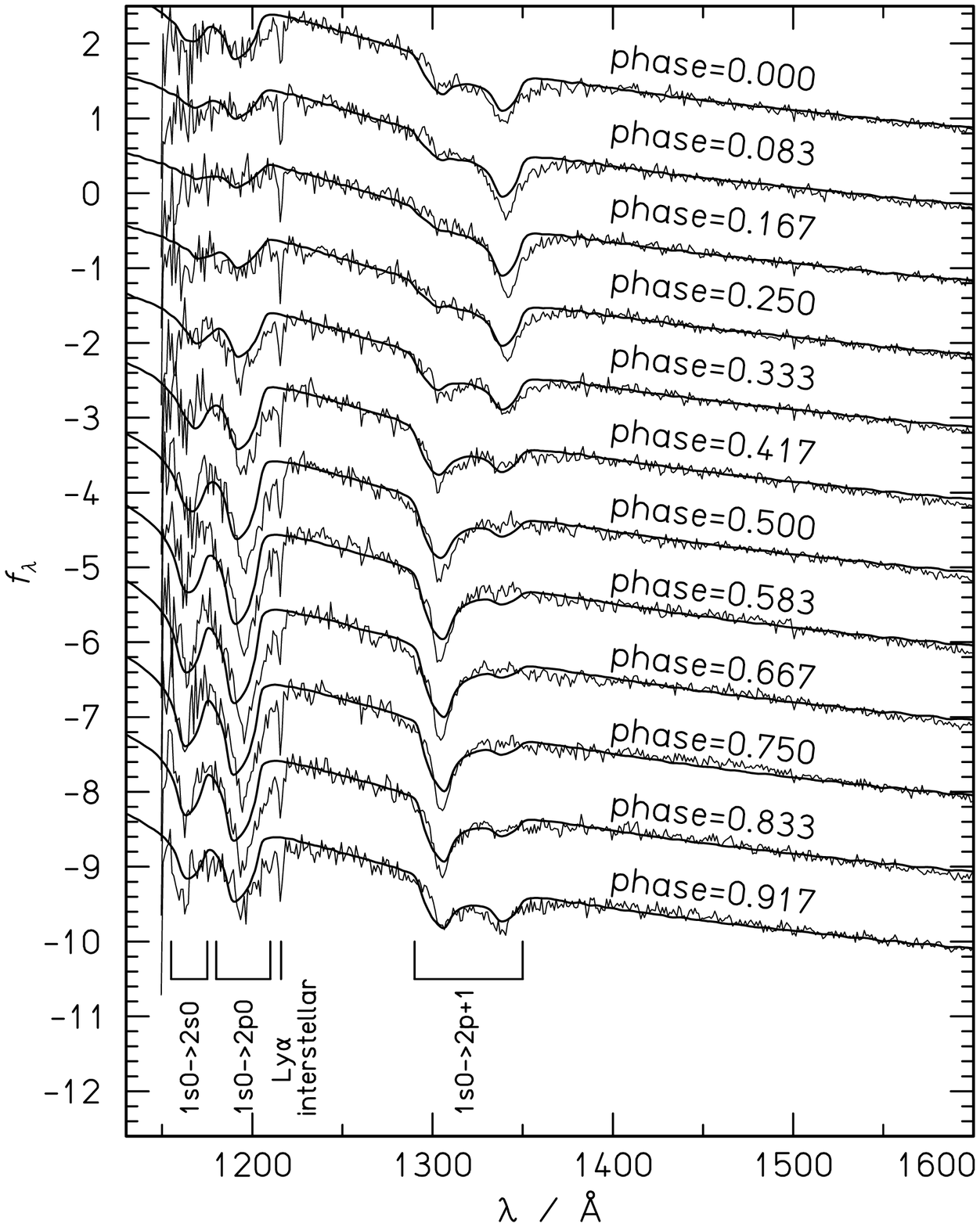]{Observed and theoretical spectra for 12
different phases across the 725 second rotation period. \label{fig1} }

\figcaption[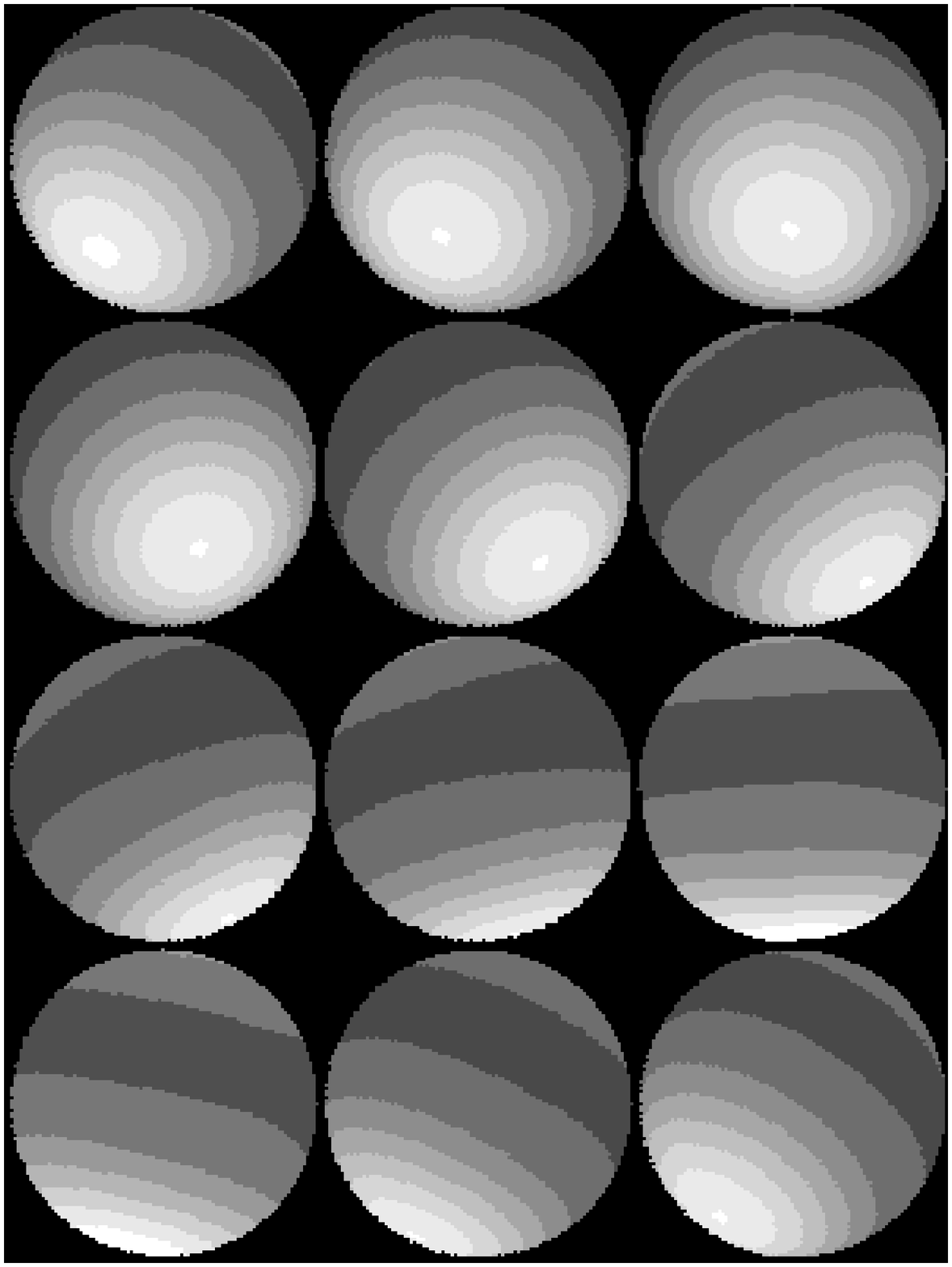]{Distribution of the magnetic field
strength over the visible hemisphere
from the best fitting model for 
the twelve phases shown in Fig.\,1. The steps in grey scale
correspond to steps of 100\,MG, the lowest field is 170\,MG (darkest
grey), 
the highest 800\,MG (white spot). An animation of this plot together with
the observed and synthetic spectra can be found on the WWW 
(http://www.astrophysik.uni-kiel.de/arbeitsgruppen/agkoe/h$-$jordan\_re0317.html)
\label{fig2}}

\figcaption[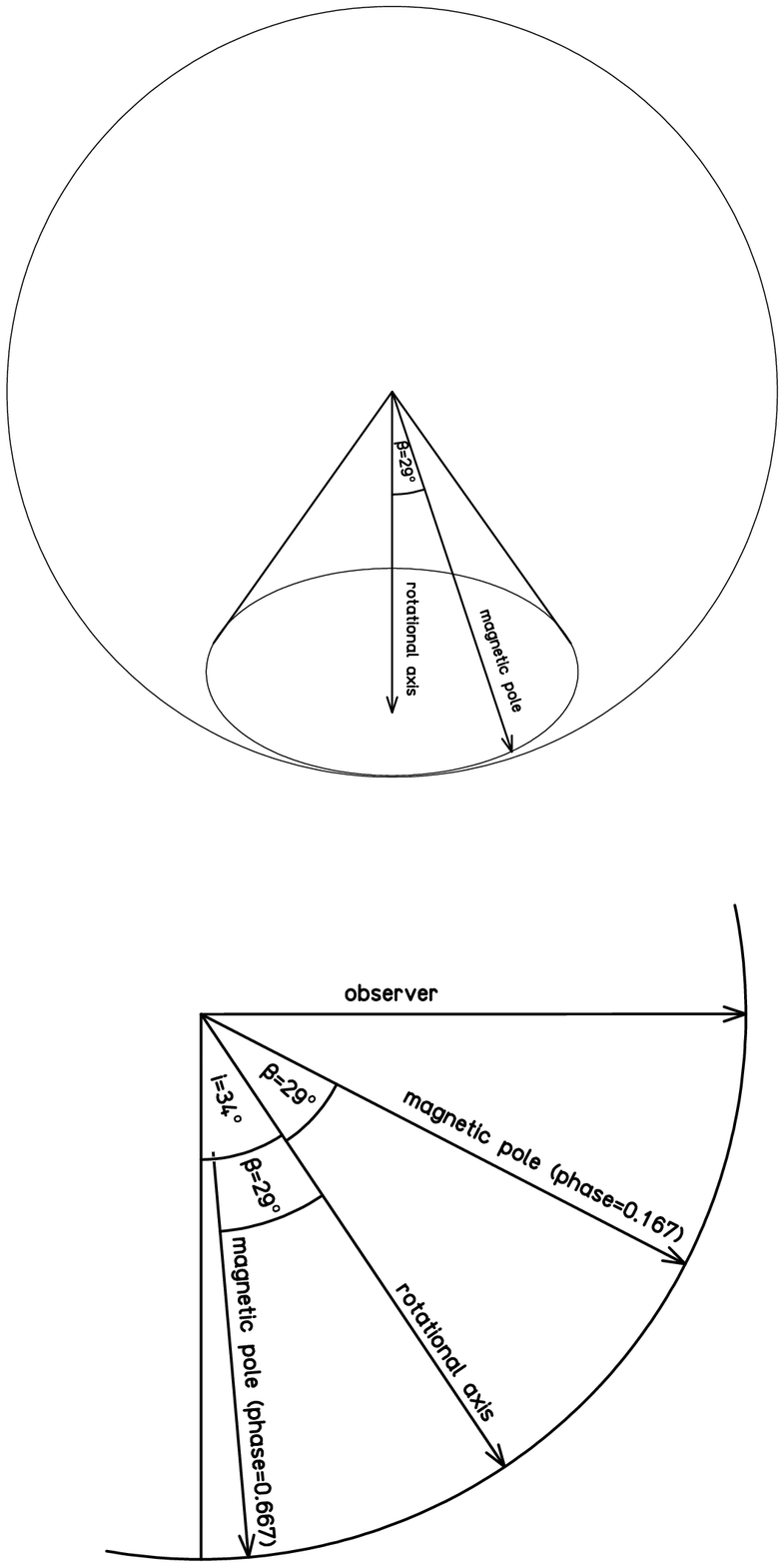]{Geometry of the best
solution model for the expansion into
spherical harmonics. The upper panel shows the projection of the
cone-like  movement of the
magnetic pole during the rotational period onto the celestial plane. 
The relative
orientations of the direction to the observer, rotational axis
and magnetic pole are shown below.\label{fig3} }

\figcaption[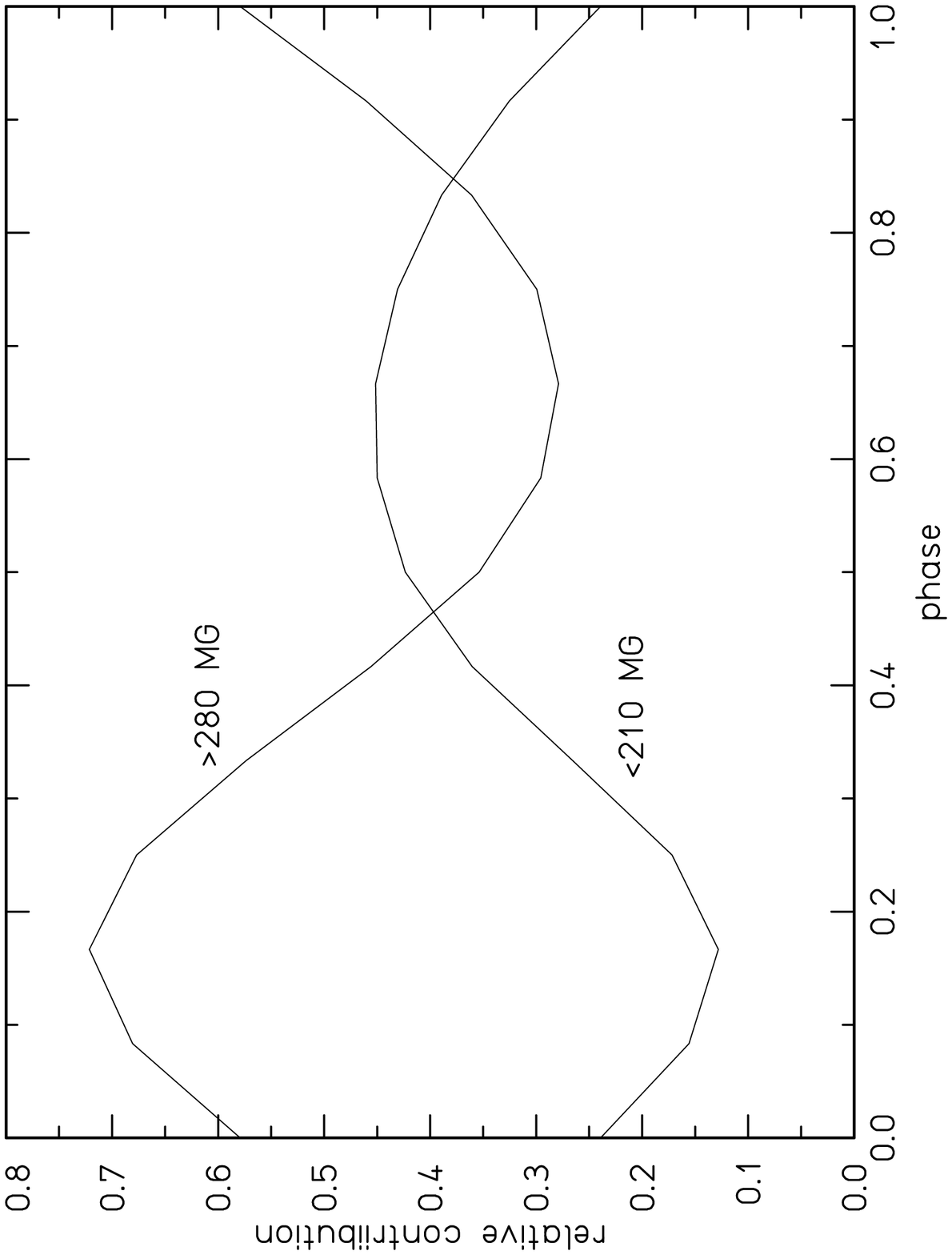]{Relative contribution of low
($<210$\,MG) and strong 
($>280$\,MG) magnetic fields  on the visible hemisphere of the model
star for the different phases. \label{fig4}}


\clearpage

\plotone{re0317_legrenge3.eps}

\clearpage

\plotone{re0317_legendre3_all2.eps}

\clearpage

\plotone{re0317_geo1_comb.eps}

\clearpage

\plotone{re0317_magfiel_hi_low.eps}

\end{document}